\documentclass[a4paper,twocolumn,11pt,unpublished]{quantumarticle}
\pdfoutput=1
\usepackage[utf8]{inputenc}
\usepackage[english]{babel}
\usepackage[T1]{fontenc}
\usepackage{amsmath}
\usepackage{hyperref}
\usepackage[numbers,sort&compress]{natbib}

\usepackage{tikz}
\usepackage{lipsum}

\usepackage{amsfonts}
\usepackage{bm}
\usepackage{qcircuit}

\newcommand{\Tr}{\textrm{Tr}}
\newcommand{\tr}{\textrm{Tr}}
\newcommand{\unit}{\mathbb{I}}

\begin{document}

\title{A scheme for direct detection of qubit-environment entanglement generated during qubit
pure dephasing}

\author{Bartosz Rzepkowski}
\affiliation{Department of Theoretical Physics, Wroc{\l}aw University of Science and Technology, 50-370 Wroc{\l}aw, Poland}

\author{Katarzyna Roszak}
\email{katarzyna.roszak@pwr.edu.pl}
\affiliation{Department of Theoretical Physics, Wroc{\l}aw University of Science and Technology, 50-370 Wroc{\l}aw, Poland}
\orcid{0000-0002-9955-4331}


\maketitle

\begin{abstract}
  We propose a scheme for the detection of qubit-environment entanglement at time $\tau$ which
  requires only operations and measurements on the qubit, all within reach of current
  experimental state-of-the-art. The scheme works for any type of
  interaction which leads to pure dephasing of the qubit as long as the initial qubit
  state is pure. It becomes particularly simple when one of the qubit states is neutral
  with respect to the environment, such as in case of the most common choice of
  the NV center spin qubit or for excitonic charge qubits, when the environment is
  initially at thermal equilibrium.
\end{abstract}

	The accessibility of entanglement in larger bipartite systems is very limited up to date, because contrarily
	to entanglement between two qubits \cite{hill97,wootters98,yu04}, the theoretical means for
	the study of such entanglement are very limited unless the joint system state is pure. The only available measure which can
	be calculated from the density matrix is Negativity \cite{vidal02,lee00a}
	or closely related logarithmic Negativity \cite{plenio05b}, the calculation of which requires diagonalization
	of a matrix of the same dimension as the joint Hilbert space of the two parties, which must be and has been
	done numerically \cite{Eisert_PRL02,Hilt_PRA09,Pernice_PRA11,salamon}. This limits the range of general conclusions which can be reached
	about the creation and behavior of entanglement. Experimentally, such entanglement
	is hardly accessible at all, since measuring Negativity would require full quantum
	state tomography, similarly as quantification of two-qubit entanglement, but as
	the technique can be done for two small systems \cite{bonk04,filipp09,liu12}, it exceeds the current experimental
	state-of-the-art once either of the potentially entangled systems becomes large.

	The problem is that the question of entanglement
	becomes important when dealing with decoherence between a
	quantum system of interest, such as a qubit, and its environment. This is because the
	presence of entanglement, although its manifestation is limited
	when straightforward qubit decoherence is of interest \cite{Helm_PRA09,Crow_PRA14}, can significantly change the effect
	that the environment has on the system in more involved procedures and algorithms,
	especially ones that involve qubit evolution post measurement \cite{viola98,viola03,calarco03,economou06,economou07,roszak15b,roszak15c,mierzejewski15}.

	The question of entanglement generation becomes more solvable once limitations on the generality of the problem
	are imposed. It has recently been shown that for evolutions which lead to pure
	dephasing of a qubit or even a larger system \cite{roszak15,roszak18a,roszak18b,chen18,chen19}, there exists a
	straightforward signature, which entanglement leaves on the state of the environment
	\cite{roszak15,roszak18a}
	(after the qubit/system state is traced out). This is, on one hand, the reason why
	it is important to know if qubit-environment entanglement (QEE) is generated, since only decoherence with generation of QEE is accompanied
	by the information about the qubit state leaking out into the environment \cite{roszak20},
	similarly as in the case of a pure environment \cite{Zurek_RMP03,Hornberger_LNP09}. On the other hand,
	it also serves as the basis for the possibility of direct measurement of QEE.

	In the following we describe a scheme for the direct experimental detection of QEE
	which occurs at time $\tau$ after the creation of a qubit superposition state.
	The scheme works only within the class of Hamiltonians that leads to qubit
	pure dephasing, but
	it involves operations and measurements performed only on the qubit.
	The required operations
	are well within reach of all systems that have been proposed as qubits,
	especially in the solid state, where pure dephasing is commonly the dominant source
	of decoherence \cite{roszak06a,Biercuk_Nature09,Bylander_NP11,Medford_PRL12,Staudacher_Science13,roszak13,Muhonen_NN14,Malinowski_PRL17,Szankowski_JPCM17}. The scheme relies
	on the fact, that although it is the state of the environment which is distinctly
	affected by the presence of QEE, it can in turn influence
	the evolution of the qubit.

	The scheme differs significantly from the one presented in Ref.~\cite{roszak19},
	as the operatons and measurements on the qubit required are different. Hence
	the scheme will be more appropriate for different qubit realizations depending
	on the experimental state-of-the-art.
	Furthermore, the scheme detects qubit-environment entanglement present
	in the state contrarily to the scheme of Ref.~\cite{roszak19},
	which serves to test if entanglement would be present at a given time, for the
	same initial state of the environment if the initial state of the qubit were a
	superpostion and not one of the pointer states which are used in the scheme.

	The scheme presented here is especially well suited for interactions
	which are asymmetric in the sense that only one of the qubit states couples to the
	environment, while the other remains unaffected by it. In this case, when the
	environment is initially at thermal equilibrium, the procedure becomes particularly
	simple, because detecting entanglement requires only the comparison of one part of
	the general procedure with regular decoherence curves (when no additional preparation
	is performed on the qubit prior to exciting a superposition state).
	Such qubit-environment systems are common; it is the case for the most common choice of
	NV center spin qubit interacting with a spin environment \cite{Kwiatkowski_PRB18,Wrachtrup_JMR16,Degen_RMP17},
	as well as for excitonic charge qubits interacting with phonons
	\cite{borri01,vagov03,vagov04,roszak06b}.
	We exemplify the working of the scheme on the latter qubit system.

	The paper is organized as follows. In Sec.~\ref{sec2} we introduce the class of
	of systems studied and the entanglement witness which is applicable for such
	systems. The proposed experimental scheme is described in Sec.~\ref{sec3}.
	The simplified measurement procedure which is applicable if the class of interactions
	is further limited is introduced in Sec.~\ref{secd}. In Sec.~\ref{sec4} we provide
	exemplary results for the workings of the scheme and Sec.~\ref{sec5}
	concludes the paper.

\section{The interaction \label{sec2}}

	We will consider a system consisting of a single qubit (Q) in the presence of an arbitrary environment (E). The most general form of a QE Hamiltonian that leads to qubit pure dephasing is given by
	\begin{equation} \label{eq:hamiltonian}
	    \hat{H} = \sum_{i=0,1} \epsilon_i |i\rangle \langle i | + \hat{H}_E + \sum_{i=0}^1 |i\rangle \langle i | \otimes \hat{V}_i.
	\end{equation}
	The first part of the Hamiltonian characterizes the qubit.
	For no processes involving energy exchange between Q and E
	to take place and therefore for the interaction to lead to pure dephasing, it must
	commute with the last, interaction, term of the Hamiltonian.
	 $\hat{H}_E$ represents the free Hamiltonian of the environment and is arbitrary.
	 The last term specifies the qubit-environment interaction with the qubit states written on the left side of the tensor product. For now we do not impose any restrictions on the components occurring in Eq.~\eqref{eq:hamiltonian}, so the interaction it describes is of most general form.

	The evolution operator of the QE system resulting from the Hamiltonian \eqref{eq:hamiltonian} can be
	formally written as
	\begin{equation}
	\label{ewolucja}
	    \hat{U}(t) = \exp\left(-\frac{i}{\hbar}\hat{H}t\right)=  \begin{bmatrix} \hat{w}_0(t) & 0 \\ 0 & \hat{w}_1(t) \end{bmatrix},
	\end{equation}
	where the matrix form is kept in terms of qubit pointer states $|0\rangle$ and $|1\rangle$,
	while the evolution of the environment is described by the operators $\hat{w}_i(t)$,
	$i=0,1$.
	These operators are given by
	\begin{equation}
	\label{w}
	    \hat{w}_i(t) = \exp \bigg (-\frac{i\epsilon_i t }{\hbar} \bigg ) \exp \bigg (-\frac{i }{\hbar} (\hat{H}_E + \hat{V}_i)t \bigg ).
	\end{equation}
	Note, that we could achieve this concise form only because the free qubit Hamiltonian
	commutes with all other Hamiltonian terms.

	\subsection{Criterion of qubit-environment entanglement}
	It has been shown in Ref.~\cite{roszak15}, that the QE system initially in a product state and undergoing evolution for time $\tau$ governed by \eqref{ewolucja} is separable, iff
	\begin{equation}
	\label{sep1}
	    \left[\hat{w}_0^\dagger(\tau)\hat{w}_1(\tau), \hat{R}(0)\right] = 0.
	\end{equation}
	Here $\hat{R}(0)$ denotes an arbitrary initial state of the environment.
	The qubits initial state has to be pure and a superposition of both pointer
	states. If we introduce the following notation
    \begin{equation}
	\label{rij}
	\hat{R}_{ij}(\tau) = \hat{w}_i(\tau) \hat{R}(0) \hat{w}_j^\dagger(\tau),
	\end{equation}
	with $i, j = 0,1$, we can reformulate the QEE criterion of Eq.~\eqref{sep1}
	to say that the QE state is entangled at time $\tau$ iff
	\begin{equation}
	\label{sep2}
	    \hat{R}_{00}(\tau) = \hat{R}_{11}(\tau).
	    \end{equation}

	\section{The scheme \label{sec3}}
	The proposed scheme for detection of QEE
	relies on the fact that the state of
	the environment influences the state of the qubit and similarly the state of the qubit
	influences the state of the environment throughout their joint evolution.
	Hence, even though the presence of QEE leaves a detectable
	mark only on the state of the environment during a simple joint evolution,
	it is possible to measure this effect when only the qubit is accessible.
	A fully indirect scheme for the detection of QEE has been recently proposed in Ref.~\cite{roszak19},
	where in fact the possibility of a given QE system to become entangled was tested, rather
	than the entanglement present in the system during the potential experiment.

	Here we show a method which allows to test the presence of entanglement at a given time $\tau$
	by further processing and measuring the state of the qubit and later comparing the post-$\tau$
	evolution of the qubit with results obtained in a test run of the same initial qubit and environment
	state.

    \subsection{Evolution of the system with intermediate measurement at time $\tau$}
    Let us assume that the initial state of the qubit is an equal superposition state,
    $|+\rangle=(|0\rangle + |1\rangle)/\sqrt{2}$, so that the QE initial state is given by
    $
        \hat{\sigma}(0) = |+\rangle\langle +| \otimes \hat{R}(0)$.
    In the first part of the scheme, we allow the two subsystems to undergo simple joint evolution as governed
    by the Hamiltonian \eqref{eq:hamiltonian} until time $\tau$,
    \begin{align}
    \label{utau}
	\hat{\sigma}(\tau) &= \hat{U}(\tau) \hat{\sigma}(0) \hat{U}^\dagger (\tau)= \frac{1}{2} \begin{bmatrix} \hat{R}_{00}(\tau) & \hat{R}_{01}(\tau) \\
	\hat{R}_{10}(\tau) & \hat{R}_{11}(\tau) \end{bmatrix}.
	\end{align}
	Time $\tau$ is singled out as the time at which we are testing QEE. In other words, a positive result
	of the proposed scheme would certify that there is entanglement in state \eqref{utau}.
	Incidentally, it is rather straightforward to generalize the proposed scheme to any initial
	qubit state, but if QEE is generated for state $|+\rangle$
	then it would also be generated for any superposition of pointer states \cite{roszak15}, $|\psi\rangle=a|0\rangle + b|1\rangle$, with $a,b\neq 0$, so there is hardly any point.

	The first step towards determining QEE in state \eqref{utau} is to measure the qubit
	in the $\{|+\rangle, |-\rangle \}$ basis, where the $|+\rangle$ state is the initial state of
	the qubit and the $|-\rangle$ state is orthogonal to it. A projective measurement
	in the qubit subspace yields the states $|\pm\rangle$ with probabilities
	\begin{equation}
	\label{prob}
	p_{\pm}(\tau) = \frac{1}{4}\  \Tr( \hat{R}_{00}(\tau) {\pm} \hat{R}_{01}(\tau) {\pm} \hat{R}_{10}(\tau) + \hat{R}_{11}(\tau) ).
	\end{equation}
	Although indirectly, the measurement also influences the environment while it leads
	to the recurrence of a product QE state, with new environmental states depending
	on the outcome of the measurement,
	\begin{equation}
    \hat{R}_{\pm}(\tau) = \frac{ \hat{R}_{00}(\tau) {\pm} \hat{R}_{01}(\tau) {\pm} \hat{R}_{10}(\tau) + \hat{R}_{11}(\tau) }{4p_{\pm}(\tau)}.
    \end{equation}
    Using $\hat{R}_{\pm}(\tau)$ we can write the post-measurement QE states at time $\tau$ as
    $
    \hat{\sigma}_{\pm}(\tau)  = |{\pm}\rangle \langle{\pm}| \otimes \hat{R}_{\pm}(\tau)$.
    Since the Hamiltonian remains unchanged
    it is straightforward to find the evolution which occurs after additional time $t$ post measurement has passed,
    \begin{align}
	&\hat{\sigma}_{\pm}(\tau + t) \\
	\nonumber
	&=\frac{1}{2} \begin{bmatrix} \hat{w}_0(t)\hat{R}_{\pm}(\tau)\hat{w}_0^\dagger(t) & {\pm}\hat{w}_0(t)\hat{R}_{\pm}(\tau)\hat{w}_1^\dagger(t) \\ {\pm}\hat{w}_1(t)\hat{R}_{\pm}(\tau)\hat{w}_0^\dagger(t) & \hat{w}_1(t)\hat{R}_{\pm}(\tau)\hat{w}_1^\dagger(t) \end{bmatrix}.
	\end{align}

    The quantity, which will later allow us to distinguish whether the pre-measurement
    state at time $\tau$ \eqref{utau} is
    entangled or separable is the evolution of the qubit coherence.
    To this end, we find the post-measurement evolution of the qubit
    by tracing out environmental degrees of freedom (the evolution of the qubit is of pure-dephasing
    type, so only off-diagonal elements of the qubit density matrix evolve)
    and the coherence is given by
    \begin{equation}
    \label{coherence}
    \rho_{\pm}^{01}(\tau+t)=\pm \frac{1}{2} \Tr\left(\hat{w}_0(t)\hat{R}_{\pm}(\tau)\hat{w}_1^\dagger(t)\right).
    \end{equation}

    The last step involves averaging the qubit coherence over the measurement outcomes, so the quantity of interest
    which, as it will turn out, contains information allowing to determine entanglement, is given
    by
    \begin{equation}
    \label{g}
    \rho_{av}^{01}(\tau+t)=p_{+}(\tau)\rho_{+}^{01}(\tau+t)-p_{-}(\tau)\rho_{-}^{01}(\tau+t).
    \end{equation}
    During the averaging, the minus sign stemming from the coherence of the initial
    qubit state $|-\rangle$ is compensated for (hence the difference and not the sum
    in the second term of Eq.~\eqref{g}).
    Experimentally this means that the same procedure, involving preparation of the initial
    qubit equal superposition state, allowing it to evolve for time $\tau$, after which a measurement of the
    qubit is performed in the $\{|\pm\rangle\}$ basis, and the following measurement of the evolution of qubit coherence, needs to be repeated a sufficient number of times, and the results have to
    be averaged regardless of the measurement outcome.
    Inserting explicit formulas for the probabilities of each measurement outcome \eqref{prob}
    and the corresponding coherences \eqref{coherence} into Eq.~\eqref{g}, we get
    a much simpler formula than the one for the coherences alone,
    \begin{equation}
    \label{14}
    \rho_{av}^{01}(\tau+t)=
    \frac{1}{4}\Tr\left[\hat{w}_0(t)\left(\hat{R}_{00}(\tau)+ \hat{R}_{11}(\tau) \right)\hat{w}_1^{\dagger}(t)\right].
    \end{equation}

	\subsection{Evolution of the comparative system}
	To determine if there is entanglement in state \eqref{utau}, the quantity \eqref{14}
	needs to be compared to the outcome of a second procedure. We will later show, that in the most
	common scenario, this second procedure reduces to a straightforward measurement of coherence
	with no additional preparation.

    Contrary to the first part, the qubit is initialized in state $|0\rangle$ with the environment
    in the same state as before,  $
        \hat{\sigma}_0(0) = |0\rangle \langle 0| \otimes \hat{R}(0)$.
    As previously, we allow the QE system to evolve jointly for time $\tau$,
    which leads to no change in the qubit state, but does lead to an evolution in the
    subspace of the environment, $
       \hat{\sigma}_0(\tau) = |0\rangle \langle 0 | \otimes \hat{R}_{00}(\tau)$,
    where $\hat{R}_{00}(\tau)$ is given by Eq.~\eqref{rij}.
    At time $\tau$ instead of conducting a measurement, the superposition $|+\rangle$
    is excited in the qubit subspace. The later QE evolution as a function of time $t$ (time elapsed
    after the excitation) differs from the undisturbed QE evolution only by the initial state of the environment.
    Here, the quantity of interest is again the coherence of the qubit as a function of time $t$
    which is obtained after tracing out the degrees of freedom of the environment
    and is given by
	\begin{equation}
	\label{coherence2}
    	\rho_{0}^{01}(\tau + t) = \frac{1}{2} \Tr( \hat{w}_0(t)\hat{R}_{00}\hat{w}_1^\dagger(t)).
	\end{equation}

	\subsection{The quantity of interest}
	To detect QEE, we need to study the difference between the average qubit coherence
	obtained from the procedure involving an intermediate measurement \eqref{14}
	and the coherence of the comparative system \eqref{coherence2},
	\begin{align}
	\label{delta}
	\Delta\rho^{01}(\tau+t)&=\rho_{av}^{01}(\tau + t)-\rho_{0}^{01}(\tau + t)\\
	\nonumber
	&= \frac{1}{4} \Tr \left[ \hat{w}_0(t) \left( \hat{R}_{11}(\tau) - \hat{R}_{00}(\tau) \right) \hat{w}_1^\dagger(t) \right].
	\end{align}
	Obviously it can be nonzero
	only if $\hat{R}_{11}(\tau) \neq \hat{R}_{00}(\tau)$, hence the quantity can be nonzero only if
	there is entanglement in state \eqref{utau}, and it is therefore an entanglement witness.
	In fact, if the difference of coherences \eqref{delta} is nonzero at any time $t$
	then there must have been QEE in the pre-measurement state at time $\tau$.

	Otherwise, either the QE state \eqref{utau} was separable and $\hat{R}_{11}(\tau) = \hat{R}_{00}(\tau)$ or the conditional environmental evolution
	operators \eqref{w} commute with one another.
	In case of commutation (the condition of their commutation  is $\left[\hat{H}_E+\hat{V}_0,\hat{H}_E+\hat{V}_1\right]=0$)
	we have
	\begin{equation}
	\Tr \left[ \hat{w}_0(t)\hat{R}_{ii}(\tau)\hat{w}_1^\dagger(t) \right]=
	\Tr \left[ \hat{w}_0(t)\hat{R}(0)  \hat{w}_1^\dagger(t) \right]
	\end{equation}
   for $i=0,1$
   and this type of entanglement cannot be detected using the scheme under study.
   In fact, the class of entangled states which will not be detected by this scheme
   are exactly the same as the class not detected by the scheme described in Ref.~\cite{roszak19}.
   For details on why schemes for detecting QEE where operations and measurements are
   restricted to the qubit subspace will not be able to detect entanglement generated
   by conditional evolution operators which commute see the Appendix.
   The protocol is illustrated in Fig.~\ref{g1}.

   \begin{figure}[h!]
		\[
		\scalebox{1.5}{\Qcircuit @C=1em @R=1em {
				\lstick{|+\rangle} & \ustick{\tau} \qw & \meter & \ustick{|-\rangle} \qw & \qw & \ustick{t} \qw & \qw \\
				\lstick{|+\rangle} & \ustick{\tau} \qw & \meter & \ustick{|+\rangle} \qw & \qw & \ustick{t} \qw & \qw \\}}
				\]
		Comparative system
		\[
		\scalebox{1.5}{\Qcircuit @C=1em @R=1em {
				\lstick{|0\rangle} & \ustick{\tau} \qw & \gate{\big ( \frac{\pi}{2} \big )} & \ustick{|+\rangle} \qw & \qw & \ustick{t} \qw & \qw \\
		}}
		\]
		\caption{Graphical representation of the general protocol for QEE detection with intermediate measurement at time $\tau$. \label{g1}}
	\end{figure}
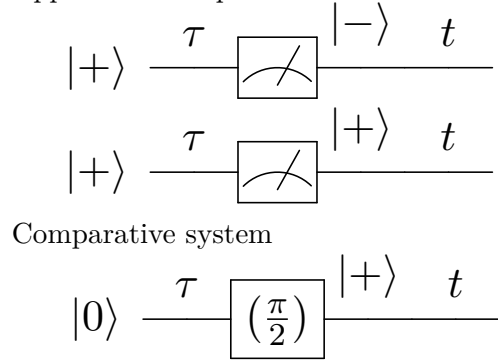


   \subsection{Asymmetric interaction \label{secd}}
   The true advantage of the scheme described here lies in the situation when the interaction
   between the qubit and its environment is asymmetric, $\hat{V}_0=\unit$,
   and the single nontrivial interaction term does not commute with the free Hamiltonian of
   the environment, $\left[\hat{H}_E,\hat{V}_1\right]\neq 0$
   (the latter condition is necessary so that the conditional evolution operators acting on the
   environment do not commute).
   This situation is reasonably common for solid state qubits, for which pure-dephasing evolutions
   are the most common source of decoherence, since it means that one of the qubit states
   does not interact with the environment. This is the case for e.~g.~ excitonic charge qubits interacting with phonons, in which case the $|0\rangle$ state consists of no exciton \cite{borri01,vagov03,vagov04,roszak06b}.
   

   If additionally the initial state of the environment is some function of the free Hamiltonian
   $\hat{H}_E$, then the whole second part of the protocol is superfluous, and a comparison of
   the results of the first part
   of the procedure
   with a straightforward
   measurement of the evolution of coherence of an initial $|+\rangle$ qubit state
   as a function of time $t$
   is enough to determine if the state \eqref{utau} is entangled.
   This is because within the specified constraints
   $
   \left[\hat{w}_0(\tau),\hat{R}(0)\right]=0$,
   since $\hat{w}_0(\tau)$
   and $\hat{R}(0)$ are functions of the same part of the Hamiltonian, namely $\hat{H}_E$.
   This means that
   $\hat{R}_{00}(\tau)=\hat{R}(0)$
   and no extra preparation time
   in the comparative evolution is necessary.

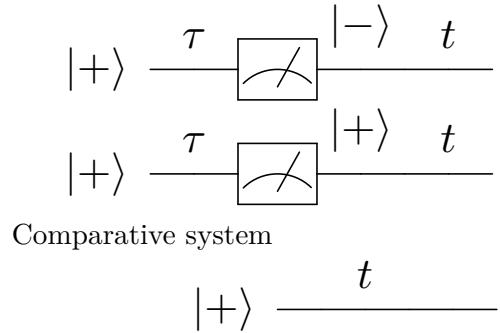
\begin{figure}[h!]
   	\[
   	\scalebox{1.5}{\Qcircuit @C=1em @R=1em {
   			\lstick{|+\rangle} & \ustick{\tau} \qw & \meter & \ustick{|-\rangle} \qw & \qw & \ustick{t} \qw & \qw \\
   			\lstick{|+\rangle} & \ustick{\tau} \qw & \meter & \ustick{|+\rangle} \qw & \qw & \ustick{t} \qw & \qw}} \]
   	Comparative system
   	\[
   	\scalebox{1.5}{\Qcircuit @C=1em @R=1em {
   			& & & \lstick{|+\rangle} & \qw & \ustick{t} \qw & \qw & \qw & \qw
   	}}
   	\]
   	\caption{Graphical representation of the simplified protocol for the system with asymmetric interaction.\label{g2}}
   \end{figure}

   The situation when the initial state of the environment is a function of the free Hamiltonian
   of the environment is the most common of all experimentally encountered
   situations, since any environment at thermal equilibrium falls into this category.
   The only situations when this does not apply, are when the environment has been specially prepared
   by prior schemes, such as dynamical polarization  \cite{london13,fisher13,wunderlich17,scheuer17,poggiali17,pagliero18,hovav18}.
   Hence, for a qubit for which one of the pointer states is neutral with respect to the
   environment, and an environment initially at thermal equilibrium, the scheme
   for entanglement detection significantly simplifies.
   In fact, it is enough to compare results of the evolution of decoherence post an intermediate
   measurement in the equal superposition basis
   averaged over the possible measurement outcomes \eqref{g} with the plain evolution of decoherence,
   and the quantity of interest \eqref{delta} simplifies to
   \begin{align}
   \label{delta2}
   \Delta\rho^{01}(\tau+t)=\rho_{av}^{01}(\tau + t)-\rho^{01}(t),
   \end{align}
   where $
   \rho^{01}(t)=\langle 0|\Tr_E\hat{\sigma}(t)|1\rangle$,
   and $\hat{\sigma}(t)$ is given by Eq.~\eqref{utau} with argument $t$ instead of $\tau$.
   The simplified protocol is illustrated in Fig.~\ref{g2}.

	\section{Example: charge qubit and phonons \label{sec4}}

	In the following, we will exemplify the validity and huge sensitivity of the procedure
	on an excitonic quantum dot qubit interacting with a bath of bulk phonons, the evolution of which is known to be entangling \cite{salamon}.
	The setup and procedure is exactly
	as in Ref.~\cite{roszak15b}.
	The qubit state $|0\rangle$ corresponds to an empty quantum dot, while qubit state $|1\rangle$
	is a ground state exciton confined in the dot, so the interaction is naturally
	asymmetric and the simplified procedure applies. The Hamiltonian of the system is given by
	\begin{equation}
	\label{ham}
	\hat{H}=\varepsilon |1\rangle\langle 1|+\sum_{\bm{k}}\hbar\omega_{\bm{k}}\hat{b}_{\bm{k}}^{\dagger}
	\hat{b}_{\bm{k}}+|1\rangle\langle 1|\otimes\sum_{\bm{k}}\left(f_{\bm{k}}^*\hat{b}_{\bm{k}}
	+f_{\bm{k}}\hat{b}_{\bm{k}}^{\dagger}\right),
	\end{equation}
	where $\varepsilon$ is the energy of the exciton, $\hbar\omega_{\bm{k}}$ are phonon energies
	corresponding to phonon creation and annihilation operators for wavevector $\bm{k}$, $\hat{b}_{\bm{k}}^{\dagger}$ and
	$\hat{b}_{\bm{k}}$.
	$f_{\bm{k}}$ in the interaction term denote the deformation
	potential coupling constants \cite{mahan00,grodecka05a}, which is the dominating interaction
	leading to exciton decoherence. They
	are given by
	\begin{equation}\label{cpl}
	f_{\bm{k}}
	=( \sigma_{\mathrm{e}} -\sigma_{\mathrm{h}} )
	\sqrt{\frac{\hbar k}{2\varrho V_{\mathrm{N}} c}}
	\int_{-\infty}^{\infty} d^3\bm{r}\psi^*(\bm{r})
	e^{-i\bm{k}\cdot\mathrm{r}}\psi(\bm{r}).
	\end{equation}
	Here $\varrho$ is the crystal density, $V_{\mathrm{N}}$ is the
	normalization volume of the phonon system, $\sigma_{e}$ and $\sigma_{h}$
	are deformation potential constants for electrons and holes,
	and $\psi(\bm{r})$ are excitonic wave functions.

	The Hamiltonian can be diagonalized exactly by means of the Weyl operator method
	\cite{andriambololona69,mahan00}.
	The exciton-phonon interaction term in the Hamiltonian
	is linear in phonon operators
	and describes a shift of the lattice equilibrium induced by the
	presence of a charge distribution in the dot. The stationary state of
	the system corresponds to the exciton being surrounded by a coherent
	cloud of phonons (representing the lattice distortion to the new
	equilibrium). The transformation that creates the coherent cloud is
	the shift
	\begin{equation}
	\hat{w}\hat{b}_{\bm{k}}\hat{w}^{\dagger}=\hat{b}_{\bm{k}}-\frac{f_{\bm{k}}}{\hbar\omega_{\bm{k}}},
	\end{equation}
	generated by the Weyl operator \cite{roszak06b}
	given by
	\begin{equation}
	\hat{w}=\exp\left[\sum_{\bm{k}}\left(
	\frac{f_{\bm{k}}}{\hbar\omega_{\bm{k}}}\hat{b}_{\bm{k}}^{\dagger}
	-\frac{f_{\bm{k}}^*}{\hbar\omega_{\bm{k}}}\hat{b}_{\bm{k}}\right)\right].
	\end{equation}
	A straightforward calculation shows that the Hamiltonian
	is diagonalized by the unitary transformation
	\begin{equation}
	\mathbb{W}=|0\rangle\langle 0|\otimes \mathbb{I}+|1\rangle\langle 1|\otimes \hat{w},
	\end{equation}
	where $\mathbb{I}$ is the identity operator and the tensor product
	refers to the carrier subsystem (first component) and its phonon
	environment (second component).

	This allows us to find the
	explicit form of conditional evolution operators acting on the environment,
	Eq.~\eqref{w},
	\begin{align}
	\label{w0}
	\hat{w}_0(t)&=e^{-\frac{i}{\hbar}\sum_{\bm{k}}\hbar\omega_{\bm{k}}\hat{b}_{\bm{k}}^{\dagger}
		\hat{b}_{\bm{k}}t},\\
	\label{w1}
	\hat{w}_1(t)&=e^{-\frac{i}{\hbar}E t}\hat{w}^{\dagger}\hat{w}_0(t)\hat{w}.
	\end{align}
	The diagonalization procedure induces a shift in the energy of the exciton, which is now given by
	\begin{equation}
	E=\varepsilon-\sum_{\bm{k}}
	\frac{|f_{\bm{k}}|^2}{\hbar\omega_{\bm{k}}}.
	\end{equation}

	\begin{figure}[tb]
		\includegraphics[width=1\columnwidth]{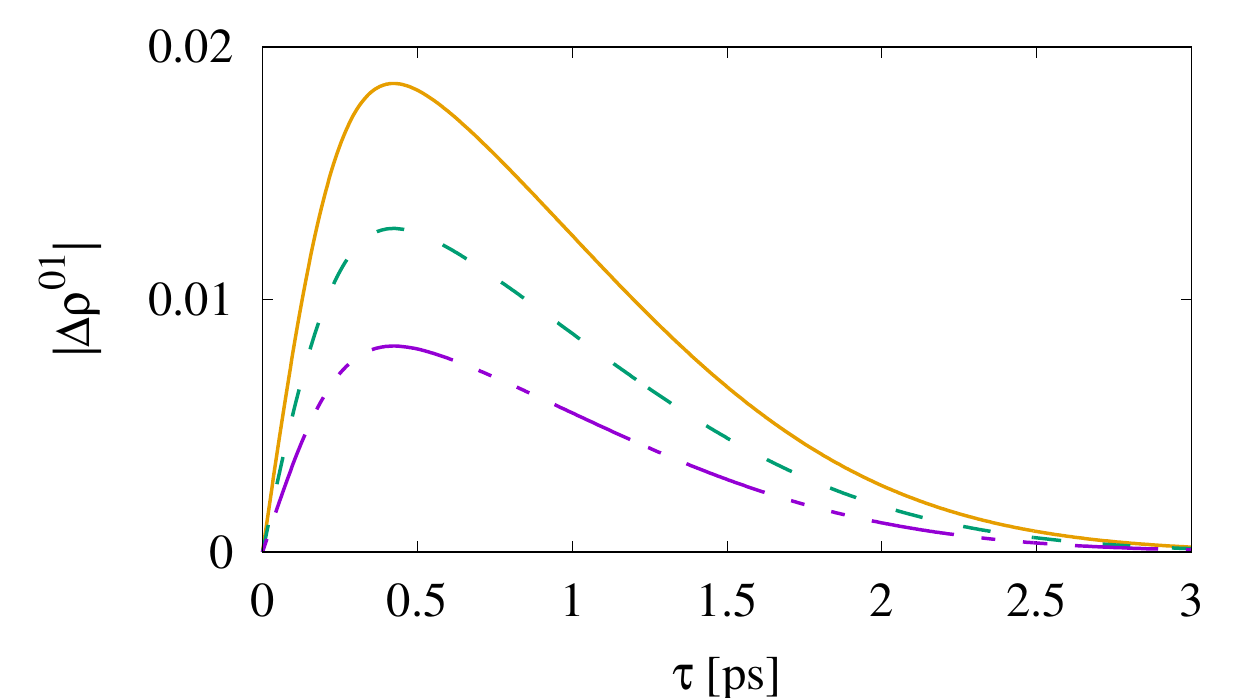}
		\caption{$\tau$ dependence of the measurable QEE witness \eqref{delta2} for a quantum dot
			excitonic qubit interacting with a phonon bath at $t\rightarrow\infty$ for
			different temperatures: $T=0$
			K - yellow solid line; $T=34$ K - green dashed line;
			$T=70$ K - purple dashed-dotted line.
		}\label{fig3}
	\end{figure}

	The explicit forms of the $\hat{w}_i(t)$ operators and the rules for multiplying Weyl
	operators \cite{roszak06b} allow us to first find the conditional density matrices
	of the environment $R_{ii}(\tau)$ using Eq.~\eqref{rij} and then the quantity of
	interest \eqref{delta2}, which is given by
	\begin{align}
	\nonumber
	\Delta\rho^{01}\left(t\right)&=\frac{1}{4}e^{-\frac{i}{\hbar}E t}
	e^{i\sum_{\bm{k}}
		\left|\frac{f_{\bm{k}}}{\hbar\omega_{\bm{k}}}\right|^2\sin\omega_{\bm{k}}t}\\
	\nonumber
	&\times e^{-\sum_{\bm{k}}
		\left|\frac{f_{\bm{k}}}{\hbar\omega_{\bm{k}}}\right|^2\left(1-\cos\omega_{\bm{k}}t\right)\left(2n_{\bm{k}}+1\right)}\\
	\label{ex}
	&\times\left[e^{2i\sum_{\bm{k}}
		\big|\frac{f_{\bm{k}}}{\hbar\omega_{\bm{k}}}\big|^2\sin\omega_{\bm{k}}\tau}-1\right].
	\end{align}
	The two phase terms dependent on time $t$ in the first line of Eq.~\eqref{ex}
	are irrelevant and can be easily eliminated by taking the absolute value. The real term
	in the next line governs the degree of decoherence and it guarantees, that no entanglement
	will be signified for infinite temperature, when the initial density matrix of the
	environment is proportional to unity and pure dephasing evolutions cannot lead to
	entanglement \cite{roszak15}. Note, that since the bath is super-Ohmic, we are dealing
	with partial pure dephasing \cite{borri01,vagov03,vagov04,roszak06b} and for long times, the degree of coherence
	stabilizes at a certain, non-zero value instead of tending to zero.
	The most important term for the detection of entanglement is given in the second
	line of Eq.~\eqref{ex}. The $\tau$ dependent phase is the signature of entanglement
	and it is similar in nature to the phase reported in Ref.~\cite{kwiatkowski20}, which signified
	the quantumness of the environment.

	Fig.~\ref{fig3} shows the $\tau$-dependence of the QEE witness given by the expression \eqref{delta2}
	in the limit $t\rightarrow\infty$ for material parameters characteristic for small GaAs quantum
	dots
	and bulk phonon modes \cite{roszak06a,roszak15b} for different temperatures.
	The presented results were obtained by modeling single particle (electron and hole, which form the exciton) wave functions $\psi(\bm{r})$ as Gaussians with $5$ nm width in the $xy$ plane and $1$ nm along $z$. The material parameters used are $\sigma_{\mathrm{e}}-\sigma_{\mathrm{h}}=9$ eV,
	$\varrho=5360$ kg/m$^{3}$, and $c=5100$ m/s.
	The shifted exciton energy is not needed (but would have been of the order of $1$ eV).

	The effect is small as to be expected,
	since the environment taken into account is very large, which yields very small amounts of
	entanglement \cite{salamon}.
	Note that contrary to the results in \cite{roszak15b}, the effect is most pronounced
	when the intermediate measurement at time $\tau$ occurs before equilibration
	(which for the studied system is after around $3$ ps).
	The quantity \eqref{delta2} is proportional to how different the conditional states of the
	environment are from each other, so it is proportional to the amount of QEE \cite{roszak20}.
	In fact, it is testimony to the extreme
	sensitivity of the method that it is visible at all.

	\section{Conclusion \label{sec5}}

	We have proposed a scheme for the detection of QEE at a certain time $\tau$
	which requires the comparison of qubit evolution after a measurement performed at said time $\tau$
	with a second, similarly simple procedure performed also only on the qubit.
	The procedure is qualitatively different from the one proposed before \cite{roszak19}
	in the fact, that it detects entanglement actually present in the system during
	its operation, when the results are compared to the ones obtained during another measurement. The characteristic feature of the procedure is that it requires an
	intermediate projective
	measurement so that information about the qubit state is transferred into the environment.
	This process is sensitive to qubit-environment entanglement, and the post-measurement
	evolution is different when the pre-measurement evolution was entangling.

	The method is applicable to any qubit-environment interaction which leads
	to qubit pure dephasing, but is particularly useful,
	if the qubit-environment interaction is asymmetric
	(so that one of the qubit pointer states does not interact with the environment)
	and the initial state of the environment is at thermal equilibrium.
	Then the procedure becomes
	particularly simple, and the post-measurement evolution needs only to be compared to the
	plain decoherence of the qubit. This means that if the evolution of qubit coherence
	post-measurement differs in any way from the evolution which would be observed
	if a superposition state was simply excited at the measurement time, then there
	must have been entanglement between the qubit and the environment pre-measurement
	at time $\tau$.

	The method is of both experimental and theoretical interest, as it requires the description only
	of qubit evolution and does not require the knowledge of the explicit state of the environment
	apart from the initial state. As an example we have studied the presence of entanglement
	between a charge qubit confined in a quantum dot with an environment of bulk phonons
	at finite temperature.
	The evolution is known to be entangling, but very weakly so; nevertheless, we have observed
	a distinct signature of qubit-environment entanglement.

	\section{Acknowledgements} \label{acknowledgements}
	K.~R.~acknowledges support from funds of the
	Polish National Science Center (NCN), Grant
	no. 2015/19/B/ST3/03152.

\onecolumn\newpage
\appendix

\section{Commuting conditional evolution operators of the environment}

In this section we consider, why QEE with commuting conditional
evolution operators of the environment
is not likely to be distinguishable by operations and measurements on the qubit alone.
The conditional operators are given in the Results section by Eq.~\eqref{w}.
The evolution is separable at time $\tau$ iff the separability criterion
\eqref{sep2} is fulfilled, which written explicitly with the help of the $\hat{w}_0(\tau)$
operators and the initial state of the environment, $\hat{R}(0)$,
takes the form
\begin{equation}
\label{sep}
\hat{w}_0(\tau)\hat{R}(0)\hat{w}_0^{\dagger}(\tau)=\hat{w}_1(\tau)\hat{R}(0)\hat{w}_1^{\dagger}(\tau);
\end{equation}
otherwise it is entangled \cite{roszak15}. We distinguish and study here
a type of entangled evolution
for which the conditional evolution operators of the environment
commute at any two times,
\begin{equation}
\label{comm}
\left[\hat{w}_0(t_0),\hat{w}_1^{\dagger}(t_1)\right]=0.
\end{equation}
Such commutation is a direct consequence of the commutation of the relevant parts of the
Hamiltonian which drive the evolution of the environment, if the qubit is respectively
in state $|0\rangle$ or $|1\rangle$,
\begin{equation}
\left[\hat{H}_E+\hat{V}_0,\hat{H}_E+\hat{V}_1\right]=0.
\end{equation}
In the following, we will call this type of entangling evolution ``weak''.

All methods that we can envision which serve to distinguish entangling from non-entangling
evolutions are based (more or less directly)
on the comparison of terms which are at most combinations (probably linear superpositions)
of terms of type
\begin{equation}
\label{r}
r(t)=\tr\left[
\hat{w}_{i_1}(t_1)\dots \hat{w}_{i_N}(t_N)\hat{R}(0)\hat{w}_{j_N}^{\dagger}(t_N)\dots w_{j_1}^{\dagger}(t_1)
\right].
\end{equation}
This is the case in the scheme proposed here as well as in the scheme of ref.~\cite{roszak19}.

If we choose two functions of the type given by Eq.~\eqref{r} which differ only by $n$
operators $\hat{w}_{i}(t)$
symmetrically on the left and right of $\hat{R}(0)$ which are closest to $\hat{R}(0)$,
these two functions must yield the same results, if the separability criterion
\eqref{sep} is fulfilled for all relevant times.
But operators under the trace can be cyclically permuted, so
\begin{equation}
\label{r2}
r(t)=\tr\left[w_{j_N}^{\dagger}(t_N)\dots w_{j_1}^{\dagger}(t_1)
w_{i_1}(t_1)\dots w_{i_N}(t_N)\hat{R}(0)
\right].
\end{equation}
If the weak entanglement criterion is fulfilled \eqref{comm} then all of the operators
corresponding to the same time which fulfill $i_k=j_k$
will cancel each other out, so the two functions \eqref{r}
which are the same for separable evolutions, will always be the same
for weakly entangled ones as well.

On the other hand, the set of pairs of functions of type \eqref{r} which are the same
for separable evolutions is a subset of the set of pairs of such functions which
fulfill the commutation criterion \eqref{comm}. Hence, it should be fairly easy to
devise a scenario which tests this commutation criterion, e.~g.~the spin echo \cite{Herzog_PR56,Gordon_PRL58}
yields perfect results only if the conditional operators commute.
This together with a QEE witness testing the separability criterion \eqref{sep} would enable to unambiguously distinguish between separable evolutions for which the conditional operators do not commute and
strongly entangled evolutions. Yet the ambiguity is still left over in the set of
evolutions for which the operators do commute, and distinguishing between such entangling
and non-entangling evolutions, although possible on measurements on the environment,
is most likely not possible by measurements on the qubit.


\begin{thebibliography}{61}
		\providecommand{\natexlab}[1]{#1}
		\providecommand{\url}[1]{\texttt{#1}}
		\expandafter\ifx\csname urlstyle\endcsname\relax
		\providecommand{\doi}[1]{doi: #1}\else
		\providecommand{\doi}{doi: \begingroup \urlstyle{rm}\Url}\fi
		
		\bibitem[Hill and Wootters(1997)]{hill97}
		Scott Hill and William~K. Wootters.
		\newblock Entanglement of a pair of quantum bits.
		\newblock \emph{Phys. Rev. Lett.}, 78:\penalty0 5022--5025, Jun 1997.
		\newblock \doi{10.1103/PhysRevLett.78.5022}.
		\newblock URL \url{https://link.aps.org/doi/10.1103/PhysRevLett.78.5022}.
		
		\bibitem[Wootters(1998)]{wootters98}
		William~K. Wootters.
		\newblock Entanglement of formation of an arbitrary state of two qubits.
		\newblock \emph{Phys. Rev. Lett.}, 80:\penalty0 2245--2248, Mar 1998.
		\newblock \doi{10.1103/PhysRevLett.80.2245}.
		\newblock URL \url{https://link.aps.org/doi/10.1103/PhysRevLett.80.2245}.
		
		\bibitem[Yu and Eberly(2004)]{yu04}
		Ting Yu and J.~H. Eberly.
		\newblock Finite-time disentanglement via spontaneous emission.
		\newblock \emph{Phys. Rev. Lett.}, 93:\penalty0 140404, Sep 2004.
		\newblock \doi{10.1103/PhysRevLett.93.140404}.
		\newblock URL \url{http://link.aps.org/doi/10.1103/PhysRevLett.93.140404}.
		
		\bibitem[Vidal and Werner(2002)]{vidal02}
		G.~Vidal and R.~F. Werner.
		\newblock Computable measure of entanglement.
		\newblock \emph{Phys. Rev. A}, 65:\penalty0 032314, Feb 2002.
		\newblock \doi{10.1103/PhysRevA.65.032314}.
		\newblock URL \url{http://link.aps.org/doi/10.1103/PhysRevA.65.032314}.
		
		\bibitem[Lee et~al.(2000)Lee, Kim, Park, and Lee]{lee00a}
		Jinhyoung Lee, M.~S. Kim, Y.~J. Park, and S.~Lee.
		\newblock Partial teleportation of entanglement in a noisy environment.
		\newblock \emph{Journal of Modern Optics}, 47\penalty0 (12):\penalty0
		2151--2164, 2000.
		\newblock \doi{10.1080/09500340008235138}.
		\newblock URL \url{https://doi.org/10.1080/09500340008235138}.
		
		\bibitem[Plenio(2005)]{plenio05b}
		M.~B. Plenio.
		\newblock Logarithmic negativity: A full entanglement monotone that is not
		convex.
		\newblock \emph{Phys. Rev. Lett.}, 95:\penalty0 090503, Aug 2005.
		\newblock \doi{10.1103/PhysRevLett.95.090503}.
		\newblock URL \url{https://link.aps.org/doi/10.1103/PhysRevLett.95.090503}.
		
		\bibitem[Eisert and Plenio(2002)]{Eisert_PRL02}
		Jens Eisert and Martin~B. Plenio.
		\newblock Quantum and classical correlations in quantum brownian motion.
		\newblock \emph{Phys. Rev. Lett.}, 89:\penalty0 137902, Sep 2002.
		\newblock \doi{10.1103/PhysRevLett.89.137902}.
		\newblock URL \url{http://link.aps.org/doi/10.1103/PhysRevLett.89.137902}.
		
		\bibitem[Hilt and Lutz(2009)]{Hilt_PRA09}
		Stefanie Hilt and Eric Lutz.
		\newblock System-bath entanglement in quantum thermodynamics.
		\newblock \emph{Phys. Rev. A}, 79:\penalty0 010101, Jan 2009.
		\newblock \doi{10.1103/PhysRevA.79.010101}.
		\newblock URL \url{http://link.aps.org/doi/10.1103/PhysRevA.79.010101}.
		
		\bibitem[Pernice and Strunz(2011)]{Pernice_PRA11}
		A.~Pernice and Walter~T. Strunz.
		\newblock Decoherence and the nature of system-environment correlations.
		\newblock \emph{Phys. Rev. A}, 84:\penalty0 062121, 2011.
		\newblock \doi{10.1103/PhysRevA.84.062121}.
		\newblock URL \url{http://link.aps.org/doi/10.1103/PhysRevA.84.062121}.
		
		\bibitem[Salamon and Roszak(2017)]{salamon}
		Tymoteusz Salamon and Katarzyna Roszak.
		\newblock Entanglement generation between a charge qubit and its bosonic
		environment during pure dephasing: Dependence on the environment size.
		\newblock \emph{Phys. Rev. A}, 96:\penalty0 032333, Sep 2017.
		\newblock \doi{10.1103/PhysRevA.96.032333}.
		\newblock URL \url{https://link.aps.org/doi/10.1103/PhysRevA.96.032333}.
		
		\bibitem[Bonk et~al.(2004)Bonk, Sarthour, deAzevedo, Bulnes, Mantovani,
		Freitas, Bonagamba, Guimar\~aes, and Oliveira]{bonk04}
		F.~A. Bonk, R.~S. Sarthour, E.~R. deAzevedo, J.~D. Bulnes, G.~L. Mantovani,
		J.~C.~C. Freitas, T.~J. Bonagamba, A.~P. Guimar\~aes, and I.~S. Oliveira.
		\newblock Quantum-state tomography for quadrupole nuclei and its application on
		a two-qubit system.
		\newblock \emph{Phys. Rev. A}, 69:\penalty0 042322, Apr 2004.
		\newblock \doi{10.1103/PhysRevA.69.042322}.
		\newblock URL \url{https://link.aps.org/doi/10.1103/PhysRevA.69.042322}.
		
		\bibitem[Filipp et~al.(2009)Filipp, Maurer, Leek, Baur, Bianchetti, Fink,
		G\"oppl, Steffen, Gambetta, Blais, and Wallraff]{filipp09}
		S.~Filipp, P.~Maurer, P.~J. Leek, M.~Baur, R.~Bianchetti, J.~M. Fink,
		M.~G\"oppl, L.~Steffen, J.~M. Gambetta, A.~Blais, and A.~Wallraff.
		\newblock Two-qubit state tomography using a joint dispersive readout.
		\newblock \emph{Phys. Rev. Lett.}, 102:\penalty0 200402, May 2009.
		\newblock \doi{10.1103/PhysRevLett.102.200402}.
		\newblock URL \url{https://link.aps.org/doi/10.1103/PhysRevLett.102.200402}.
		
		\bibitem[Liu et~al.(2012)Liu, Zhang, Liu, Chen, and Yuan]{liu12}
		Wei-Tao Liu, Ting Zhang, Ji-Ying Liu, Ping-Xing Chen, and Jian-Min Yuan.
		\newblock Experimental quantum state tomography via compressed sampling.
		\newblock \emph{Phys. Rev. Lett.}, 108:\penalty0 170403, Apr 2012.
		\newblock \doi{10.1103/PhysRevLett.108.170403}.
		\newblock URL \url{https://link.aps.org/doi/10.1103/PhysRevLett.108.170403}.
		
		\bibitem[Helm and Strunz(2009)]{Helm_PRA09}
		Julius Helm and Walter~T. Strunz.
		\newblock Quantum decoherence of two qubits.
		\newblock \emph{Phys. Rev. A}, 80:\penalty0 042108, 2009.
		\newblock \doi{10.1103/PhysRevA.80.042108}.
		\newblock URL \url{http://link.aps.org/doi/10.1103/PhysRevA.80.042108}.
		
		\bibitem[Crow and Joynt(2014)]{Crow_PRA14}
		Daniel Crow and Robert Joynt.
		\newblock Classical simulation of quantum dephasing and depolarizing noise.
		\newblock \emph{Phys. Rev. A}, 89:\penalty0 042123, Apr 2014.
		\newblock \doi{10.1103/PhysRevA.89.042123}.
		\newblock URL \url{http://link.aps.org/doi/10.1103/PhysRevA.89.042123}.
		
		\bibitem[Viola and Lloyd(1998)]{viola98}
		Lorenza Viola and Seth Lloyd.
		\newblock Dynamical suppression of decoherence in two-state quantum systems.
		\newblock \emph{Phys. Rev. A}, 58:\penalty0 2733--2744, Oct 1998.
		\newblock \doi{10.1103/PhysRevA.58.2733}.
		\newblock URL \url{https://link.aps.org/doi/10.1103/PhysRevA.58.2733}.
		
		\bibitem[Viola and Knill(2003)]{viola03}
		Lorenza Viola and Emanuel Knill.
		\newblock Robust dynamical decoupling of quantum systems with bounded controls.
		\newblock \emph{Phys. Rev. Lett.}, 90:\penalty0 037901, Jan 2003.
		\newblock \doi{10.1103/PhysRevLett.90.037901}.
		\newblock URL \url{https://link.aps.org/doi/10.1103/PhysRevLett.90.037901}.
		
		\bibitem[Calarco et~al.(2003)Calarco, Datta, Fedichev, Pazy, and
		Zoller]{calarco03}
		T.~Calarco, A.~Datta, P.~Fedichev, E.~Pazy, and P.~Zoller.
		\newblock Spin-based all-optical quantum computation with quantum dots:
		Understanding and suppressing decoherence.
		\newblock \emph{Phys. Rev. A}, 68:\penalty0 012310, Jul 2003.
		\newblock \doi{10.1103/PhysRevA.68.012310}.
		\newblock URL \url{https://link.aps.org/doi/10.1103/PhysRevA.68.012310}.
		
		\bibitem[Economou et~al.(2006)Economou, Sham, Wu, and Steel]{economou06}
		Sophia~E. Economou, L.~J. Sham, Yanwen Wu, and D.~G. Steel.
		\newblock Proposal for optical u(1) rotations of electron spin trapped in a
		quantum dot.
		\newblock \emph{Phys. Rev. B}, 74:\penalty0 205415, Nov 2006.
		\newblock \doi{10.1103/PhysRevB.74.205415}.
		\newblock URL \url{https://link.aps.org/doi/10.1103/PhysRevB.74.205415}.
		
		\bibitem[Economou and Reinecke(2007)]{economou07}
		Sophia~E. Economou and T.~L. Reinecke.
		\newblock Theory of fast optical spin rotation in a quantum dot based on
		geometric phases and trapped states.
		\newblock \emph{Phys. Rev. Lett.}, 99:\penalty0 217401, Nov 2007.
		\newblock \doi{10.1103/PhysRevLett.99.217401}.
		\newblock URL \url{https://link.aps.org/doi/10.1103/PhysRevLett.99.217401}.
		
		\bibitem[Roszak et~al.(2015{\natexlab{a}})Roszak, Filip, and
		Novotn{\'y}]{roszak15b}
		Katarzyna Roszak, Radim Filip, and Tom{\'a\v s} Novotn{\'y}.
		\newblock Decoherence control by quantum decoherence itself.
		\newblock \emph{Sci. Rep}, 5:\penalty0 9796, 2015{\natexlab{a}}.
		\newblock \doi{https://doi.org/10.1038/srep09796}.
		
		\bibitem[Roszak et~al.(2015{\natexlab{b}})Roszak, Marcinowski, and
		Machnikowski]{roszak15c}
		Katarzyna Roszak, \L{}ukasz Marcinowski, and Pawe\l{} Machnikowski.
		\newblock Decoherence-enhanced quantum measurement of a quantum-dot spin qubit.
		\newblock \emph{Phys. Rev. A}, 91:\penalty0 032118, Mar 2015{\natexlab{b}}.
		\newblock \doi{10.1103/PhysRevA.91.032118}.
		\newblock URL \url{https://link.aps.org/doi/10.1103/PhysRevA.91.032118}.
		
		\bibitem[Mierzejewski et~al.(2015)Mierzejewski, Bon\ifmmode~\check{c}\else
		\v{c}\fi{}a, and Dajka]{mierzejewski15}
		Marcin Mierzejewski, Janez Bon\ifmmode~\check{c}\else \v{c}\fi{}a, and Jerzy
		Dajka.
		\newblock Reversal of relaxation due to a dephasing environment.
		\newblock \emph{Phys. Rev. A}, 91:\penalty0 052112, May 2015.
		\newblock \doi{10.1103/PhysRevA.91.052112}.
		\newblock URL \url{https://link.aps.org/doi/10.1103/PhysRevA.91.052112}.
		
		\bibitem[Roszak and Cywi\ifmmode~\acute{n}\else \'{n}\fi{}ski(2015)]{roszak15}
		Katarzyna Roszak and \L{}ukasz Cywi\ifmmode~\acute{n}\else \'{n}\fi{}ski.
		\newblock Characterization and measurement of qubit-environment-entanglement
		generation during pure dephasing.
		\newblock \emph{Phys. Rev. A}, 92:\penalty0 032310, Sep 2015.
		\newblock \doi{10.1103/PhysRevA.92.032310}.
		\newblock URL \url{https://link.aps.org/doi/10.1103/PhysRevA.92.032310}.
		
		\bibitem[Roszak(2018)]{roszak18a}
		Katarzyna Roszak.
		\newblock Criteria for system-environment entanglement generation for systems
		of any size in pure-dephasing evolutions.
		\newblock \emph{Phys. Rev. A}, 98:\penalty0 052344, Nov 2018.
		\newblock \doi{10.1103/PhysRevA.98.052344}.
		\newblock URL \url{https://link.aps.org/doi/10.1103/PhysRevA.98.052344}.
		
		\bibitem[Roszak and Cywi\ifmmode~\acute{n}\else \'{n}\fi{}ski(2018)]{roszak18b}
		Katarzyna Roszak and \L{}ukasz Cywi\ifmmode~\acute{n}\else \'{n}\fi{}ski.
		\newblock Equivalence of qubit-environment entanglement and discord generation
		via pure dephasing interactions and the resulting consequences.
		\newblock \emph{Phys. Rev. A}, 97:\penalty0 012306, Jan 2018.
		\newblock \doi{10.1103/PhysRevA.97.012306}.
		\newblock URL \url{https://link.aps.org/doi/10.1103/PhysRevA.97.012306}.
		
		\bibitem[Chen et~al.(2018)Chen, Gneiting, Lo, Chen, and Nori]{chen18}
		Hong-Bin Chen, Clemens Gneiting, Ping-Yuan Lo, Yueh-Nan Chen, and Franco Nori.
		\newblock Simulating open quantum systems with hamiltonian ensembles and the
		nonclassicality of the dynamics.
		\newblock \emph{Phys. Rev. Lett.}, 120:\penalty0 030403, Jan 2018.
		\newblock \doi{10.1103/PhysRevLett.120.030403}.
		\newblock URL \url{https://link.aps.org/doi/10.1103/PhysRevLett.120.030403}.
		
		\bibitem[Chen et~al.(2019)Chen, Lo, Gneiting, Bae, Chen, and Nori]{chen19}
		Hong-Bin Chen, Ping-Yuan Lo, Clemens Gneiting, Joonwoo Bae, Yueh-Nan Chen, and
		Franco Nori.
		\newblock Quantifying the nonclassicality of pure dephasing.
		\newblock \emph{Nature Comm.}, 10:\penalty0 3794, 2019.
		\newblock \doi{10.1038/s41467-019-11502-4}.
		\newblock URL \url{https://doi.org/10.1038/s41467-019-11502-4}.
		
		\bibitem[Roszak(2019)]{roszak20}
		Katarzyna Roszak.
		\newblock A measure of qubit environment entanglement for pure dephasing
		evolutions.
		\newblock 2019.
		\newblock URL \url{https://arxiv.org/abs/1912.07317}.
		\newblock quant-ph/1912.07317.
		
		\bibitem[Zurek(2003)]{Zurek_RMP03}
		Wojciech~Hubert Zurek.
		\newblock Decoherence, einselection, and the quantum origins of the classical.
		\newblock \emph{Rev. Mod. Phys.}, 75:\penalty0 715--775, May 2003.
		\newblock \doi{10.1103/RevModPhys.75.715}.
		\newblock URL \url{https://link.aps.org/doi/10.1103/RevModPhys.75.715}.
		
		\bibitem[Hornberger(2009)]{Hornberger_LNP09}
		Klaus Hornberger.
		\newblock Introduction to decoherence theory.
		\newblock \emph{Lect.~Notes Phys.}, 768:\penalty0 221, 2009.
		\newblock \doi{10.1007/978-3-540-88169-8_5}.
		
		\bibitem[Roszak and Machnikowski(2006{\natexlab{a}})]{roszak06a}
		Katarzyna Roszak and Pawe\l{} Machnikowski.
		\newblock Complete disentanglement by partial pure dephasing.
		\newblock \emph{Phys. Rev. A}, 73:\penalty0 022313, Feb 2006{\natexlab{a}}.
		\newblock \doi{10.1103/PhysRevA.73.022313}.
		\newblock URL \url{https://link.aps.org/doi/10.1103/PhysRevA.73.022313}.
		
		\bibitem[Biercuk et~al.(2009)Biercuk, Uys, VanDevender, Shiga, Itano, and
		Bollinger]{Biercuk_Nature09}
		Michael~J. Biercuk, Hermann Uys, Aaron~P. VanDevender, Nobuyasu Shiga, Wayne~M.
		Itano, and John~J. Bollinger.
		\newblock Optimized dynamical decoupling in a model quantum memory.
		\newblock \emph{Nature}, 458:\penalty0 996, 2009.
		\newblock \doi{10.1038/nature07951}.
		
		\bibitem[Bylander et~al.(2011)Bylander, Gustavsson, Yan, Yoshihara, Harrabi,
		Fitch, Cory, Nakamura, Tsai, and Oliver]{Bylander_NP11}
		Jonas Bylander, Simon Gustavsson, Fei Yan, Fumiki Yoshihara, Khalil Harrabi,
		George Fitch, David~G. Cory, Yasunobu Nakamura, Jaw-Shen Tsai, and William~D.
		Oliver.
		\newblock Dynamical decoupling and noise spectroscopy with a superconducting
		flux qubit.
		\newblock \emph{Nat.~Phys.}, 7:\penalty0 565, 2011.
		\newblock \doi{10.1038/nphys1994}.
		
		\bibitem[Medford et~al.(2012)Medford, Cywi\'{n}ski, Barthel, Marcus, Hanson,
		and Gossard]{Medford_PRL12}
		J.~Medford, {\L}.~Cywi\'{n}ski, C.~Barthel, C.~M. Marcus, M.~P. Hanson, and
		A.~C. Gossard.
		\newblock Scaling of dynamical decoupling for spin qubits.
		\newblock \emph{Phys.\ Rev.\ Lett.}, 108:\penalty0 086802, 2012.
		\newblock \doi{10.1103/PhysRevLett.108.086802}.
		
		\bibitem[Staudacher et~al.(2013)Staudacher, Shi, Pezzagna, Meijer, Du, Meriles,
		Reinhard, and Wrachtrup]{Staudacher_Science13}
		T.~Staudacher, F.~Shi, S.~Pezzagna, J.~Meijer, J.~Du, C.~A. Meriles,
		F.~Reinhard, and J.~Wrachtrup.
		\newblock Nuclear magnetic resonance spectroscopy on a (5-nanometer)$^{3}$
		sample volume.
		\newblock \emph{Science}, 339:\penalty0 561, 2013.
		\newblock \doi{10.1126/science.1231675}.
		
		\bibitem[Roszak et~al.(2013)Roszak, Mazurek, and Horodecki]{roszak13}
		K.~Roszak, P.~Mazurek, and P.~Horodecki.
		\newblock Anomalous decay of quantum correlations of quantum dot qubits.
		\newblock \emph{Phys. Rev. A}, 87:\penalty0 062308, 2013.
		\newblock \doi{10.1103/PhysRevA.87.062308}.
		
		\bibitem[Muhonen et~al.(2014)Muhonen, Dehollain, Laucht, Hudson, Kalra,
		Sekiguchi, Itoh, Jamieson, McCallum, Dzurak, and Morello]{Muhonen_NN14}
		Juha~T. Muhonen, Juan~P. Dehollain, Arne Laucht, Fay~E. Hudson, Rachpon Kalra,
		Takeharu Sekiguchi, Kohei~M. Itoh, David~N. Jamieson, Jeffrey~C. McCallum,
		Andrew~S. Dzurak, and Andrea Morello.
		\newblock Storing quantum information for 30 seconds in a nanoelectronic
		device.
		\newblock \emph{Nature Nanotechnology}, 9:\penalty0 986, 2014.
		\newblock \doi{10.1038/nnano.2014.211}.
		
		\bibitem[Malinowski et~al.(2017)Malinowski, Martins, Cywi{\'n}ski, Rudner,
		Nissen, Fallahi, Gardner, Manfra, Marcus, and Kuemmeth]{Malinowski_PRL17}
		Filip~K. Malinowski, Frederico Martins, {\L}ukasz Cywi{\'n}ski, Mark~S. Rudner,
		Peter~D. Nissen, Saeed Fallahi, Geoffrey~C. Gardner, Michael~J. Manfra,
		Charles~M. Marcus, and Ferdinand Kuemmeth.
		\newblock Spectrum of the nuclear environment for gaas spin qubits.
		\newblock \emph{Phys.\ Rev.\ Lett.}, 118:\penalty0 177702, 2017.
		\newblock \doi{10.1103/PhysRevLett.118.177702}.
		
		\bibitem[Sza\'nkowski et~al.(2017)Sza\'nkowski, Ramon, Krzywda, Kwiatkowski,
		and Cywi\'nski]{Szankowski_JPCM17}
		P.~Sza\'nkowski, G.~Ramon, J.~Krzywda, D.~Kwiatkowski, and \L. Cywi\'nski.
		\newblock Environmental noise spectroscopy with qubits subjected to dynamical
		decoupling.
		\newblock \emph{J. Phys.:Condens. Matter}, 29\penalty0 (33):\penalty0 333001,
		2017.
		\newblock \doi{10.1088/1361-648X/aa7648}.
		
		\bibitem[Roszak et~al.(2019)Roszak, Kwiatkowski, and
		Cywi\ifmmode~\acute{n}\else \'{n}\fi{}ski]{roszak19}
		Katarzyna Roszak, Damian Kwiatkowski, and \L{}ukasz Cywi\ifmmode~\acute{n}\else
		\'{n}\fi{}ski.
		\newblock How to detect qubit-environment entanglement generated during qubit
		dephasing.
		\newblock \emph{Phys. Rev. A}, 100:\penalty0 022318, Aug 2019.
		\newblock \doi{10.1103/PhysRevA.100.022318}.
		\newblock URL \url{https://link.aps.org/doi/10.1103/PhysRevA.100.022318}.
		
		\bibitem[Kwiatkowski and Cywi\ifmmode~\acute{n}\else
		\'{n}\fi{}ski(2018)]{Kwiatkowski_PRB18}
		Damian Kwiatkowski and \L{}ukasz Cywi\ifmmode~\acute{n}\else \'{n}\fi{}ski.
		\newblock Decoherence of two entangled spin qubits coupled to an interacting
		sparse nuclear spin bath: Application to nitrogen vacancy centers.
		\newblock \emph{Phys. Rev. B}, 98:\penalty0 155202, Oct 2018.
		\newblock \doi{10.1103/PhysRevB.98.155202}.
		\newblock URL \url{https://link.aps.org/doi/10.1103/PhysRevB.98.155202}.
		
		\bibitem[Wrachtrup and Finkler(2016)]{Wrachtrup_JMR16}
		J.~Wrachtrup and A.~Finkler.
		\newblock Single spin magnetic resonance.
		\newblock \emph{J.~Magn.~Res.}, 369:\penalty0 225, 2016.
		\newblock \doi{10.1016/j.jmr.2016.06.017}.
		
		\bibitem[Degen et~al.(2017)Degen, Reinhard, and Cappellaro]{Degen_RMP17}
		C.~L. Degen, F.~Reinhard, and P.~Cappellaro.
		\newblock Quantum sensing.
		\newblock \emph{Rev. Mod. Phys.}, 89:\penalty0 035002, Jul 2017.
		\newblock \doi{10.1103/RevModPhys.89.035002}.
		
		\bibitem[Borri et~al.(2001)Borri, Langbein, Schneider, Woggon, Sellin, Ouyang,
		and Bimberg]{borri01}
		P.~Borri, W.~Langbein, S.~Schneider, U.~Woggon, R.~L. Sellin, D.~Ouyang, and
		D.~Bimberg.
		\newblock Ultralong dephasing time in ingaas quantum dots.
		\newblock \emph{Phys. Rev. Lett.}, 87:\penalty0 157401, Sep 2001.
		\newblock \doi{10.1103/PhysRevLett.87.157401}.
		\newblock URL \url{https://link.aps.org/doi/10.1103/PhysRevLett.87.157401}.
		
		\bibitem[Vagov et~al.(2003)Vagov, Axt, and Kuhn]{vagov03}
		A.~Vagov, V.~M. Axt, and T.~Kuhn.
		\newblock Impact of pure dephasing on the nonlinear optical response of single
		quantum dots and dot ensembles.
		\newblock \emph{Phys. Rev. B}, 67:\penalty0 115338, Mar 2003.
		\newblock \doi{10.1103/PhysRevB.67.115338}.
		\newblock URL \url{https://link.aps.org/doi/10.1103/PhysRevB.67.115338}.
		
		\bibitem[Vagov et~al.(2004)Vagov, Axt, Kuhn, Langbein, Borri, and
		Woggon]{vagov04}
		A.~Vagov, V.~M. Axt, T.~Kuhn, W.~Langbein, P.~Borri, and U.~Woggon.
		\newblock Nonmonotonous temperature dependence of the initial decoherence in
		quantum dots.
		\newblock \emph{Phys. Rev. B}, 70:\penalty0 201305, Nov 2004.
		\newblock \doi{10.1103/PhysRevB.70.201305}.
		\newblock URL \url{https://link.aps.org/doi/10.1103/PhysRevB.70.201305}.
		
		\bibitem[Roszak and Machnikowski(2006{\natexlab{b}})]{roszak06b}
		K.~Roszak and P.~Machnikowski.
		\newblock {``Which path''} decoherence in quantum dot experiments.
		\newblock \emph{Phys. Lett. A}, 351\penalty0 (4-5):\penalty0 251--256,
		2006{\natexlab{b}}.
		\newblock \doi{https://doi.org/10.1016/j.physleta.2005.11.012}.
		
		\bibitem[London et~al.(2013)London, Scheuer, Cai, Schwarz, Retzker, Plenio,
		Katagiri, Teraji, Koizumi, Isoya, Fischer, McGuinness, Naydenov, and
		Jelezko]{london13}
		P.~London, J.~Scheuer, J.-M. Cai, I.~Schwarz, A.~Retzker, M.~B. Plenio,
		M.~Katagiri, T.~Teraji, S.~Koizumi, J.~Isoya, R.~Fischer, L.~P. McGuinness,
		B.~Naydenov, and F.~Jelezko.
		\newblock Detecting and polarizing nuclear spins with double resonance on a
		single electron spin.
		\newblock \emph{Phys. Rev. Lett.}, 111:\penalty0 067601, Aug 2013.
		\newblock \doi{10.1103/PhysRevLett.111.067601}.
		\newblock URL \url{https://link.aps.org/doi/10.1103/PhysRevLett.111.067601}.
		
		\bibitem[Fischer et~al.(2013)Fischer, Bretschneider, London, Budker, Gershoni,
		and Frydman]{fisher13}
		Ran Fischer, Christian~O. Bretschneider, Paz London, Dmitry Budker, David
		Gershoni, and Lucio Frydman.
		\newblock Bulk nuclear polarization enhanced at room temperature by optical
		pumping.
		\newblock \emph{Phys. Rev. Lett.}, 111:\penalty0 057601, Jul 2013.
		\newblock \doi{10.1103/PhysRevLett.111.057601}.
		\newblock URL \url{https://link.aps.org/doi/10.1103/PhysRevLett.111.057601}.
		
		\bibitem[Wunderlich et~al.(2017)Wunderlich, Kohlrautz, Abel, Haase, and
		Meijer]{wunderlich17}
		Ralf Wunderlich, Jonas Kohlrautz, Bernd Abel, J\"urgen Haase, and Jan Meijer.
		\newblock Optically induced cross relaxation via nitrogen-related defects for
		bulk diamond $^{13}\mathrm{C}$ hyperpolarization.
		\newblock \emph{Phys. Rev. B}, 96:\penalty0 220407, Dec 2017.
		\newblock \doi{10.1103/PhysRevB.96.220407}.
		\newblock URL \url{https://link.aps.org/doi/10.1103/PhysRevB.96.220407}.
		
		\bibitem[Scheuer et~al.(2017)Scheuer, Schwartz, M\"uller, Chen, Dhand, Plenio,
		Naydenov, and Jelezko]{scheuer17}
		Jochen Scheuer, Ilai Schwartz, Samuel M\"uller, Qiong Chen, Ish Dhand,
		Martin~B. Plenio, Boris Naydenov, and Fedor Jelezko.
		\newblock Robust techniques for polarization and detection of nuclear spin
		ensembles.
		\newblock \emph{Phys. Rev. B}, 96:\penalty0 174436, Nov 2017.
		\newblock \doi{10.1103/PhysRevB.96.174436}.
		\newblock URL \url{https://link.aps.org/doi/10.1103/PhysRevB.96.174436}.
		
		\bibitem[Poggiali et~al.(2017)Poggiali, Cappellaro, and Fabbri]{poggiali17}
		F.~Poggiali, P.~Cappellaro, and N.~Fabbri.
		\newblock Measurement of the excited-state transverse hyperfine coupling in nv
		centers via dynamic nuclear polarization.
		\newblock \emph{Phys. Rev. B}, 95:\penalty0 195308, May 2017.
		\newblock \doi{10.1103/PhysRevB.95.195308}.
		\newblock URL \url{https://link.aps.org/doi/10.1103/PhysRevB.95.195308}.
		
		\bibitem[Pagliero et~al.(2018)Pagliero, Rao, Zangara, Dhomkar, Wong, Abril,
		Aslam, Parker, King, Avalos, Ajoy, Wrachtrup, Pines, and Meriles]{pagliero18}
		Daniela Pagliero, K.~R.~Koteswara Rao, Pablo~R. Zangara, Siddharth Dhomkar,
		Henry~H. Wong, Andrea Abril, Nabeel Aslam, Anna Parker, Jonathan King,
		Claudia~E. Avalos, Ashok Ajoy, Joerg Wrachtrup, Alexander Pines, and
		Carlos~A. Meriles.
		\newblock Multispin-assisted optical pumping of bulk $^{13}\mathrm{C}$ nuclear
		spin polarization in diamond.
		\newblock \emph{Phys. Rev. B}, 97:\penalty0 024422, Jan 2018.
		\newblock \doi{10.1103/PhysRevB.97.024422}.
		\newblock URL \url{https://link.aps.org/doi/10.1103/PhysRevB.97.024422}.
		
		\bibitem[Hovav et~al.(2018)Hovav, Naydenov, Jelezko, and Bar-Gill]{hovav18}
		Y.~Hovav, B.~Naydenov, F.~Jelezko, and N.~Bar-Gill.
		\newblock Low-field nuclear polarization using nitrogen vacancy centers in
		diamonds.
		\newblock \emph{Phys. Rev. Lett.}, 120:\penalty0 060405, Feb 2018.
		\newblock \doi{10.1103/PhysRevLett.120.060405}.
		\newblock URL \url{https://link.aps.org/doi/10.1103/PhysRevLett.120.060405}.
		
		\bibitem[Mahan(2000)]{mahan00}
		Gerald~D. Mahan.
		\newblock \emph{Many-Particle Physics}.
		\newblock Kluwer, New York, 2000.
		\newblock \doi{10.1007/978-1-4757-5714-9}.
		
		\bibitem[Grodecka et~al.(2005)Grodecka, Jacak, Machnikowski, and
		Roszak]{grodecka05a}
		A.~Grodecka, L.~Jacak, P.~Machnikowski, and K.~Roszak.
		\newblock Phonon impact on the coherent control of quantum states in
		semiconductor quantum dots.
		\newblock In \emph{Quantum Dots: Research Developments}, page~47. Nova Science,
		NY, 2005.
		\newblock ISBN 9781594544064.
		\newblock URL \url{https://arxiv.org/abs/cond-mat/0404364}.
		\newblock cond-mat/0404364.
		
		\bibitem[Andriambololona(1969)]{andriambololona69}
		Raoelina Andriambololona.
		\newblock {A Diagonalization Method of a Hamiltonian in Second-Quantization.
			I}.
		\newblock \emph{Progress of Theoretical Physics}, 41\penalty0 (4):\penalty0
		1064--1080, 04 1969.
		\newblock ISSN 0033-068X.
		\newblock \doi{10.1143/PTP.41.1064}.
		\newblock URL \url{https://doi.org/10.1143/PTP.41.1064}.
		
		\bibitem[Kwiatkowski et~al.(2020)Kwiatkowski, Sza\ifmmode~\acute{n}\else
		\'{n}\fi{}kowski, and Cywi\ifmmode~\acute{n}\else
		\'{n}\fi{}ski]{kwiatkowski20}
		Damian Kwiatkowski, Piotr Sza\ifmmode~\acute{n}\else \'{n}\fi{}kowski, and
		\L{}ukasz Cywi\ifmmode~\acute{n}\else \'{n}\fi{}ski.
		\newblock Influence of nuclear spin polarization on the spin-echo signal of an
		nv-center qubit.
		\newblock \emph{Phys. Rev. B}, 101:\penalty0 155412, Apr 2020.
		\newblock \doi{10.1103/PhysRevB.101.155412}.
		\newblock URL \url{https://link.aps.org/doi/10.1103/PhysRevB.101.155412}.
		
		\bibitem[Herzog and Hahn(1956)]{Herzog_PR56}
		B.~Herzog and E.~L. Hahn.
		\newblock Transient nuclear induction and double nuclear resonance in solids.
		\newblock \emph{Phys. Rev.}, 103:\penalty0 148--166, Jul 1956.
		\newblock \doi{10.1103/PhysRev.103.148}.
		\newblock URL \url{http://link.aps.org/doi/10.1103/PhysRev.103.148}.
		
		\bibitem[Gordon and Bowers(1958)]{Gordon_PRL58}
		J.~P. Gordon and K.~D. Bowers.
		\newblock Microwave spin echoes from donor electrons in silicon.
		\newblock \emph{Phys. Rev. Lett.}, 1:\penalty0 368--370, Nov 1958.
		\newblock \doi{10.1103/PhysRevLett.1.368}.
		\newblock URL \url{http://link.aps.org/doi/10.1103/PhysRevLett.1.368}.
		
	\end{thebibliography}
\end{document}